\documentclass[epj,twocolumn]{webofc}
\usepackage[varg]{txfonts}   
\usepackage{multirow}
\woctitle{Hadron Collider Physics symposium 2012}

\begin{document}
\newcommand{\Bs}{$\mathrm{B_{\mathrm s}^0}$~}
\newcommand{\Bzero}{$\mathrm{B^0}$~}
\newcommand{\Tomumu}{$\rightarrow \mu^+ \mu^-$}
%
%
\title{Studies of rare B hadron decays to leptons at hadron colliders}
%
%

\author{Vincenzo Chiochia\inst{1}\fnsep\thanks{\email{vincenzo.chiochia@cern.ch}} 
}

\institute{Physik-Institut, University of Zurich, Winterthurerstr. 190, 8057 Zurich, Switzerland}

\abstract{
Rare B hadron decays provide an excellent test bench for the Standard Model and can probe new physics models.
We review the experimental progress of the searches for rare leptonic B decays ($b\rightarrow \ell^+ \ell^-$ and $b\rightarrow s \ell^+ \ell^-$)  at LHC and Tevatron experiments.
}
\maketitle
\section{Introduction\label{sec:Introduction}}
B hadron decays mediated by flavor-changing neutral currents (FCNC) are suppressed in the Standard Model (SM) and are sensitive to new particle contributions through loop diagrams. Searches for B$^0_{\mathrm{s,d}}$ meson decays to dimuons were performed both at the Tevatron and the LHC, where the large samples of B hadrons provide a unique opportunity to study rare decays.  In addition, the rates and angular distributions of $b \rightarrow s \mu^+ \mu^-$ decay products are also sensitive to new physics.
In Section~\ref{sec:raredecays} we review the experimental status of the searches for the \Bs and \Bzero rare decays to dimuons at hadron colliders. In Section~\ref{sec:b2smm} the angular analysis and measurements of branching ratios for various $b \rightarrow s \mu^+ \mu^-$ decays performed by the CDF experiment is summarised.
%
%
\section{Searches for the rare decays $\mathrm{B_{s,d}^0}\rightarrow \mu^+\mu^-$\label{sec:raredecays}}
The rare decays \Bs\Tomumu and \Bzero\Tomumu are highly suppressed in the SM of particle physics due to their FCNC nature. These decays are forbidden at tree level and can proceed only through higher order diagrams, such as electroweak penguin and box diagrams. Furthermore, the decays are helicity suppressed and require an internal quark annihilation within the B meson. The SM predictions for the \Bs and \Bzero branching ratios to dimuons are $(3.23 \pm 0.27) \times 10^{-9}$ and $(1.07 \pm 0.10) \times 10^{-10}$, respectively, with the main uncertainty resulting from the value of the B meson decay constant $f_B$~\cite{Buras:2012ru}. The comparison with experimental results requires the inclusion of soft-photon radiation and \Bs-$\overline{\mathrm{B^0}}_\mathrm{s}$ oscillations effects each yielding O(10\%) corrections to the predicted decay rate. Several extensions of the SM, such as supersymmetric models and models with a non-standard Higgs sector, predict enhancements or suppressions to the branching fractions for these rare decays. The rare nature of the processes and the rather precise SM predictions make these decays an excellent probe for physics beyond the SM.

At the LHC searches for the \Bs and \Bzero decays to dimuons were performed by the ATLAS, CMS and LHCb experiments using datasets of $pp$ collisions at center-of-mass energies $\sqrt{s}=7$ or 8~TeV collected between 2010 and 2012.
The CMS experiment performed a simultaneous search for the rare decays \Bs$\rightarrow \mu^+ \mu^-$ and \Bzero$\rightarrow \mu^+ \mu^-$ using an integrated luminosity of 5 fb$^{-1}$ collected at the center-of-mass energy $\sqrt{s}=7$~TeV~\cite{Chatrchyan:2012rga}. An event-counting experiment was performed in dimuon mass regions around the \Bs and \Bzero masses and all selection criteria were established before observing the signal region. A normalization sample of events with $\mathrm{B^+}\rightarrow J/\psi \mathrm{K^+}$ decays (where $J/\psi \rightarrow \mu^+\mu^-$) was used to remove uncertainties related to the b-quark production cross section and the integrated luminosity. Combinatorial backgrounds were evaluated from the data in dimuon invariant mass sidebands while backgrounds from B decays were assessed with Monte Carlo (MC) simulation. The analysis was performed separately in two channels, {\it barrel} and {\it endcap}, and then combined for the final result. The barrel channel contained the candidates where both muons have $|\eta|<1.4$ and the endcap channel included those where at least one muon had $|\eta|>1.4$. B\Tomumu candidates with traverse momentum above 6.5(8.5)~GeV in the barrel(endcap) were formed by two oppositely-charged muons  originating from a common vertex and with an invariant mass in the range $4.9 < m_{\mu\mu} < 5.9$~GeV. The (sub)leading muon transverse momentum was required to be larger than (4.0)4.5~GeV in the barrel  and (4.2)4.5~GeV in the endcap. Further selection cuts were applied on the B candidate isolation variables, decay length significance and three-dimensional pointing angle.
\begin{figure*}[htbp]
\begin{center}
\includegraphics[width=10cm]{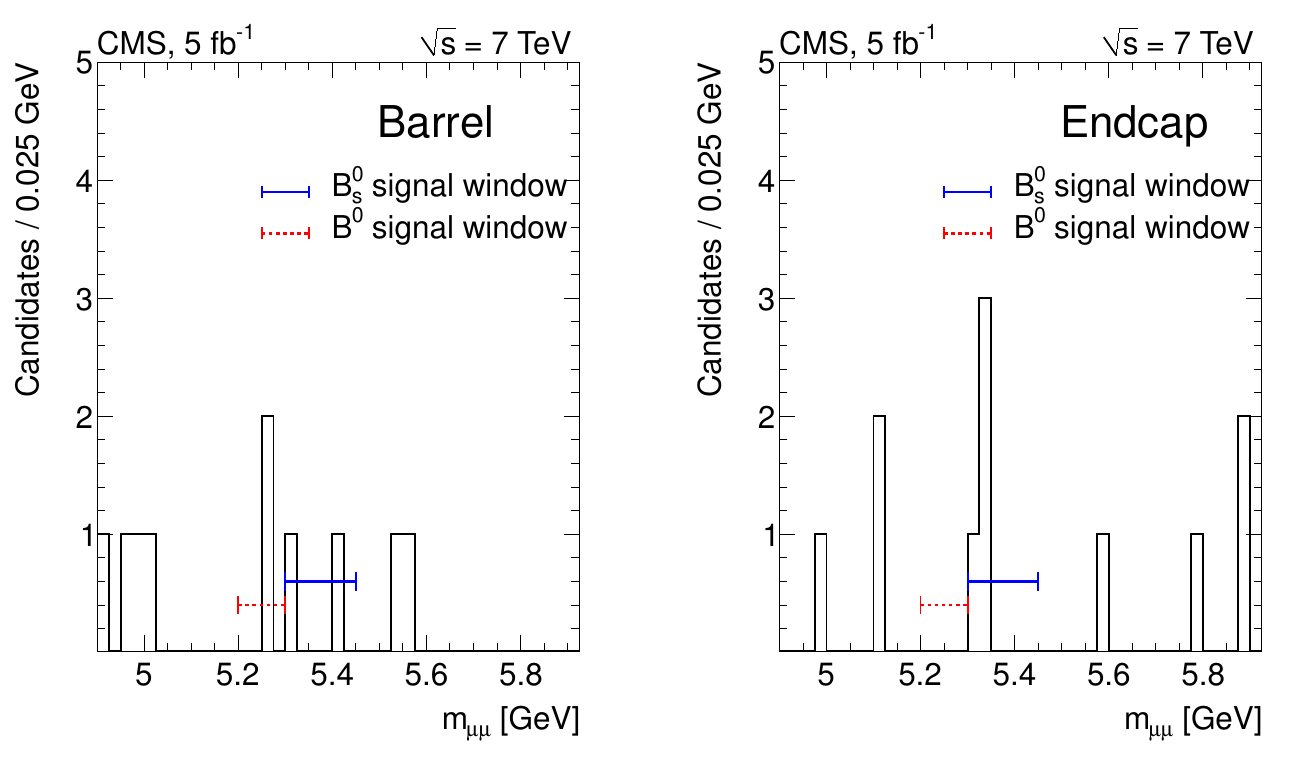}
\caption{Dimuon invariant-mass distributions measured by CMS in the barrel (left) and endcap (right) channels. The signal windows for \Bs and \Bzero are indicated by horizontal lines}
\label{fig:CMSdimuonmass}
\end{center}
\end{figure*}

The branching fraction was measured using the following equation
\begin{equation}
{\cal B}(\mathrm{B}\rightarrow \mu^+ \mu^-) = \frac{N_{sig}}{N_{norm}} \frac{f_u}{f_s} \frac{\epsilon_{norm}}{\epsilon_{sig}}{{\cal B}(\mathrm{B^+}\rightarrow J/\psi \mathrm{K^+}),}\label{eq:branching}
\end{equation}
where $\epsilon_{sig}(\epsilon_{norm})$ is the total signal(normalization) efficiency, $N_{norm}$ is the number of reconstructed $\mathrm{B^+}\rightarrow J/\psi \mathrm{K^+}$ decays, ${\cal B}({\mathrm B^+})$ is the branching fraction for the normalization channel, $f_u/f_s$ is the ratio of the ${\mathrm B^+}$ and \Bs production cross sections, and $N_{sig}$ is the background-subtracted number of observed candidates in the signal window. Figure~\ref{fig:CMSdimuonmass} shows the measured dimuon invariant-mass distributions. Six events were observed in the \Bs\Tomumu signal windows, while two events were observed in the \Bzero\Tomumu channel. This observation is consistent with the SM expectation for signal plus background. Exclusion limits on the branching fractions were obtained with the CL$_\mathrm{s}$ method and are reported in Table~\ref{tab:bsmmlimits}.
\begin{table*}[htdp]
\caption{Expected and observed upper limits (95\% CL) for the B$_{\mathrm{s,d}}$ branching fractions to dimuons.}
\begin{center}
\begin{tabular}{c|cc|cc|c}
\multirow{2}{*}{Experiment} & \multicolumn{2}{c|}{${\cal B}$(\Bs\Tomumu)} & \multicolumn{2}{c|}{${\cal B}$(\Bzero\Tomumu)} & \multirow{2}{*}{Ref.}\\ 
   &    Exp. & Obs. & Exp. & Obs. & \\ \hline
ATLAS & $23\times 10^{-9}$ & $22 \times 10^{-9}$ & - & - & \cite{Aad:2012pn}\\
CMS	& $8.4\times 10^{-9}$ & $7.7\times 10^{-9}$ &  $1.6\times 10^{-9}$ & $1.8\times 10^{-9}$ & \cite{Chatrchyan:2012rga} \\
LHCb & $7.2\times 10^{-9}$ & $4.5\times 10^{-9}$ & $1.1\times 10^{-9}$ & $1.0\times 10^{-9}$ & \cite{Aaij:2012ac,Aaij:2011rja}\\ \hline
LHC combined & $6.1\times 10^{-9}$ & $4.2\times 10^{-9}$ & $7.3\times 10^{-10}$ & $8.1\times 10^{-10}$  & \cite{CMS:2012rva} \\ \hline
CDF & $13\times 10^{-9}$ & $31\times 10^{-9}$ & $4.2\times 10^{-9}$ & $4.6\times 10^{-9}$ & \cite{CDFbsmmJan2013}\\
D0  & $23\times 10^{-9}$ & $15\times 10^{-9}$ & - & - & \cite{D0bsmm}\\
\end{tabular}
\end{center}
\label{tab:bsmmlimits}
\end{table*}%

\begin{figure}[htbp]
\begin{center}
\includegraphics[width=7cm]{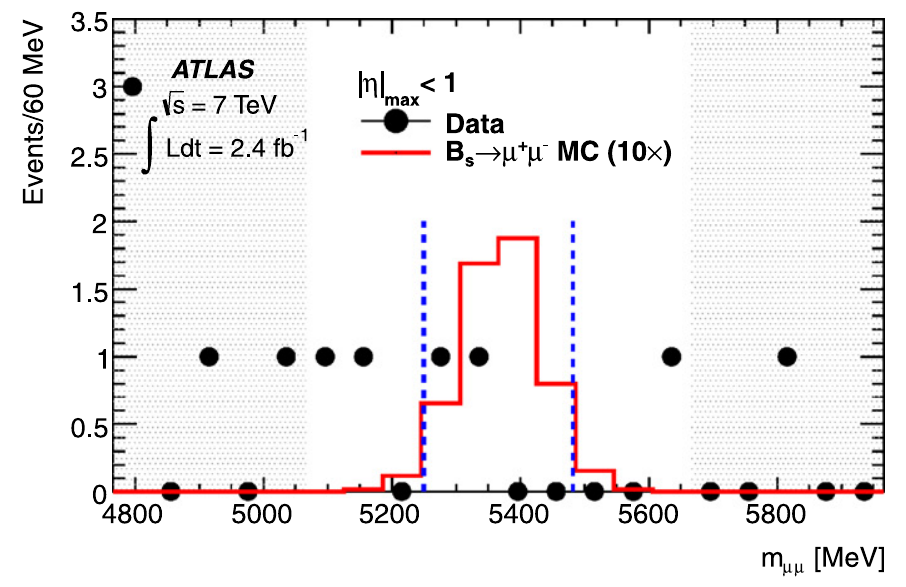}
\caption{Invariant-mass distribution of dimuon candidates in ATLAS data for the pseudorapidity region with the best invariant-mass resolution. The invariant-mass distribution for the selected candidates in data (dots), the signal (continuous line) as predicted by MC assuming ${\cal B}$(\Bs$\rightarrow \mu^+\mu^-$)=$3.5\times10^{-8}$, and two dashed vertical lines corresponding to the optimized signal region. The grey areas correspond to the sidebands used in the analysis.}
\label{fig:atlasdimuonmass}
\end{center}
\end{figure}

The ATLAS experiment performed a search for the rare decay \Bs\Tomumu using 2.4~fb$^{-1}$ collected at $\sqrt{s}=7$~TeV~\cite{Aad:2012pn}. The sensitivity to the \Bzero\Tomumu decay is beyond the reach of the current analysis. Hence only a limit on the \Bs decay was derived by assuming the \Bzero branching ratio to be negligible. The di-muon mass region $5.066 < m_{\mu\mu} < 5.666$~GeV was removed from the analysis until the procedures for event selection, signal and limit extractions were fully defined. The sample of signal candidates was selected with a multivariate classifier, trained on a fraction of the events from the di-muon invariant-mass sidebands. 14 discriminating variables were used in the selection, including the isolation, decay angle and proper decay length significance, vertex separation, impact parameter of the decay products as well as the B hadron and single muon transverse momenta. Three regions of different mass resolution and hence signal-to-background ratio were defined and selection cuts were optimized independently. The three categories were defined by the intervals $|\eta_{max}|=0-1$, $1-1.5$ and $1.5-2.5$, where $\eta_{max}$ is the largest pseudorapidity value of the two muons in each event. The width of the search region (116 to 171~MeV) and the classifier output threshold (0.234 to 0.270) were optimised following the prescriptions of Ref.~\cite{Punzi:2003bu}. The MC validation for the kinematic distributions and classifier output was performed on the normalization channel $\mathrm{B^+}\rightarrow J/\psi \mathrm{K^+}$ and residual discrepancies between data and simulation were treated as systematic uncertainties. The independence of the classifier output on the dimuon mass was tested by re-training the same classifier in an higher-mass unblinded region. The branching ratio was obtained with Eq.~\ref{eq:branching}, where the acceptance and selection efficiencies were extracted from MC simulation. In each mass-resolution category the \Bs\Tomumu signal yield was obtained from the number of events observed in the search window (see Figure~\ref{fig:atlasdimuonmass}), the number of background events in the side-bands  (excluding the events used in the classifier training and cut optimisation processes), and the small amount of resonant background. The expected and observed upper limits on the branching ratio at 95\% confidence level (CL) were obtained with the CL$_\mathrm{s}$ method and are reported in Table~\ref{tab:bsmmlimits}.

The exclusion limits on the branching ratios for \Bs and \Bzero decays to dimuons obtained from the ATLAS, CMS and LHCb searches were combined using the CL$_\mathrm{s}$ method~\cite{CMS:2012rva}. The LHCb results used in the combination were obtained analyzing integrated luminosities of 1~fb$^{-1}$ and 0.037~fb$^{-1}$ collected in 2011 and 2010, respectively, at $\sqrt{s}=7$~TeV~\cite{Aaij:2012ac,Aaij:2011rja}. No ATLAS data were used in the limit combination for the \Bzero decay. The results of the LHC combination are reported in Table~\ref{tab:bsmmlimits} and constitute the most stringent exclusions limits to date for both decays. 

The LHCb experiment recently updated the search combining datasets corresponding to integrated luminosities of 1.1~fb$^{-1}$ and 1.0~fb$^{-1}$ collected at $\sqrt{s}=8$~TeV and 7~TeV, respectively~\cite{:2012ct}. The analysis yielded an excess of events in the \Bs\Tomumu channel with respect to the background expectation, with a signal significance of 3.5 standard deviations.  An unbinned maximum-likelihood fit gave a branching fraction  ${\cal B}$(\Bs\Tomumu)$=(3.2^{+1.5}_{-1.2})\times 10^{-9}$, where the statistical uncertainty is 95\% of the total uncertainty. This value is in agreement with the SM expectation. The number of candidates in the \Bzero\Tomumu mass window is consistent with the background expectation and the observed(expected) 95\% CL exclusion limit is $9.4(7.1)\times 10^{-10}$. A more detailed discussion of the LHCb results can be found in these proceedings~\cite{LHCb-HCP2012}.

At the Tevatron similar searches were performed by the CDF and D0 experiments. The most stringent constraint is provided by the D0 collaboration, that has fully analyzed the Run II $p\bar{p}$ dataset corresponding to an integrated luminosity of 10.4~fb$^{-1}$~\cite{D0bsmm}. The mass resolution is not sufficient to separate the \Bs and \Bzero decays and a search for \Bs decays was performed assuming no contribution from \Bzero. Decay candidates were identified by selecting two high-quality muons of opposite charge forming a three-dimensional vertex well separated from the primary interaction. Backgrounds  in the low mass region are dominated by sequential decays $b\rightarrow c \mu X\rightarrow \mu \mu X'$ while on the high mass region double semileptonic decays are dominant. From the peaking backgrounds the \Bs$\rightarrow$KK is the dominant contribution. Oppositely charged muons with $p_T>1.5$~GeV were selected and a multivariate discriminant was used to differentiate between signal and backgrounds. Two boosted decision trees (BDT) were trained against the two background categories using 30 variables. The optimal BDT output cuts were determined by optimizing the expected limit the \Bs branching ratio and were set to 0.19 and 0.26. The expected number of events in the signal and control regions was determined by applying a log likelihood fit to the dimuon mass distribution using an exponential plus constant functional form. Three events in the signal region were observed after unblinding, which is consistent with expected background. The observed and expected upper limits improve substantially the previous D0 result and are given in Table~\ref{tab:bsmmlimits}.

The CDF collaboration has analyzed the full Run II data sample~\cite{CDFbsmmJan2013}, with an event selection based on a neural network using 14 input variables. It required high quality muon candidates with transverse momentum $p_T > 2.0 (2.2)$~GeV in the central (forward) region. The muon pairs were required to have an invariant mass in the range $4.669 < m_{\mu\mu} < 5.969$~GeV and are constrained to originate from a common well measured three-dimensional vertex. A likelihood method together with a selection based on the track energy loss were used to further suppress contributions from hadrons misidentified as muons. The combinatoric background was estimated by fitting a fixed slope first order polynomial to the mass sidebands while the peaking background from B$\rightarrow h h'$ decays is evaluated from MC, where the hadron misidentification probabilities were extracted from data. The number of candidates in the \Bzero mass window after unblinding are consistent with the background expectations and the corresponding exclusions limits are reported in Table~\ref{tab:bsmmlimits}. The search in the \Bs window yielded an excess of events in the region of neural network output $\nu > 0.97$ with a background-only p-value of 0.94\%. The combination of all neural network bins has a p-value of 6.8\% with respect to a background plus SM-signal hypothesis. 
The corresponding observed and expected exclusion limits obtained with the CL$_\mathrm{s}$ method are given in Table~\ref{tab:bsmmlimits}.

\begin{figure}[htbp]
\begin{center}
\includegraphics[width=\columnwidth]{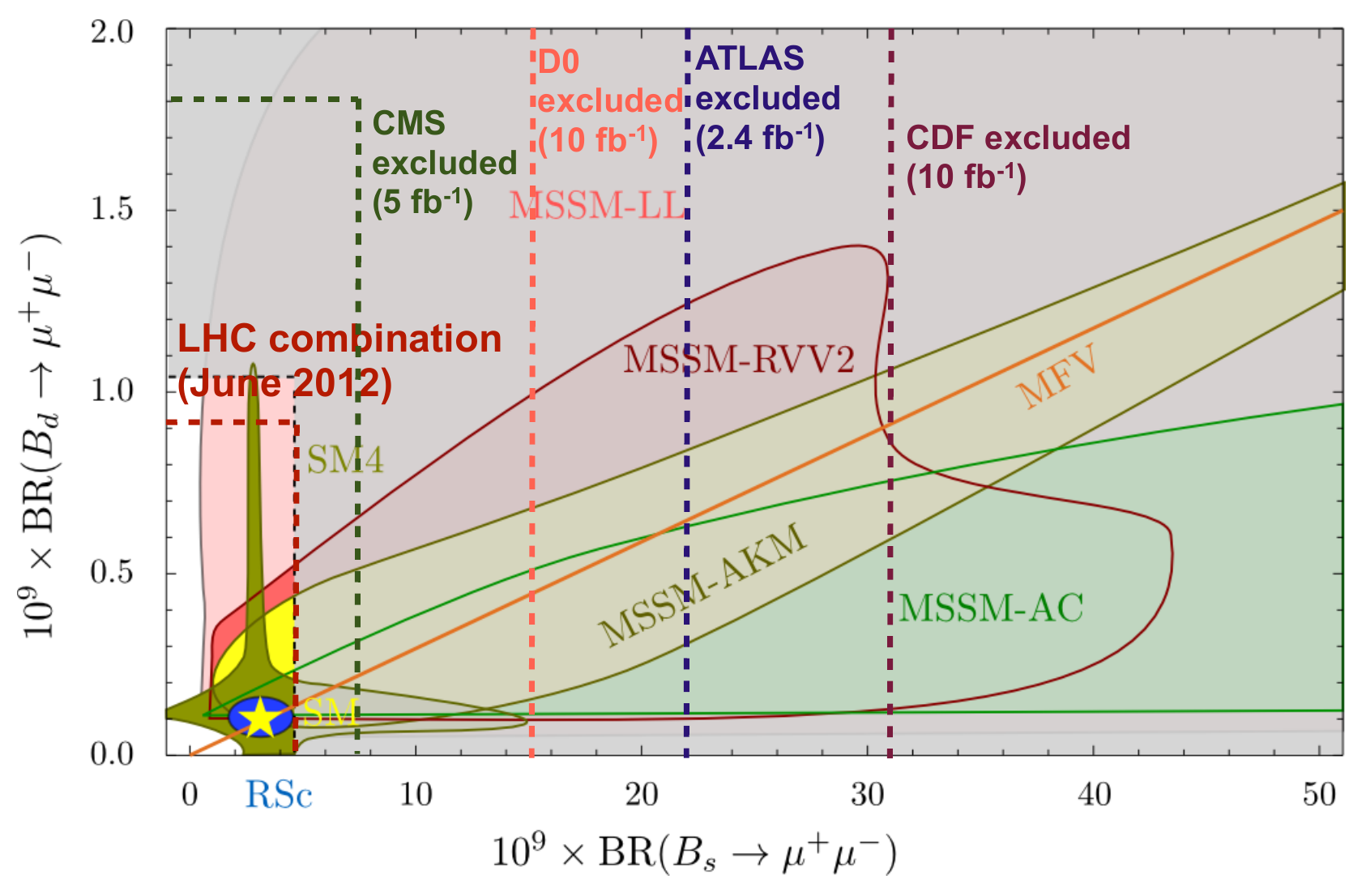}
\caption{Summary of the measured exclusion limits at 95\% CL for the \Bs and \Bzero decays to dimuons at LHC and Tevatron. Areas on the right hand side of the dashed line are excluded. The regions predicted by SM4 and four SUSY models are represented by the colored contours. The SM expectation is marked by a star (from~\cite{Straub:2012jb}).}
\label{fig:NPimpact}
\end{center}
\end{figure}

The impact of the $\mathrm{B_{s,d}}$\Tomumu searches at hadron colliders on new physics models is represented in Figure~\ref{fig:NPimpact}, showing the correlation between \Bs and \Bzero branching ratio to dimuons for models with Minimal Flavor Violation (MFV), the SM with a sequential fourth generation (SM4) and four SUSY flavour models (MSSM-RVV2/AKM/AC/LL)~\cite{Straub:2012jb}. A large part of the parameter space of the supersymmetric models, where $\tan\beta$ can be large, is ruled out by the current constraints. While the recent LHCb observation indicates a value of ${\cal B}$(\Bs\Tomumu) consistent with the SM, large enhancements of ${\cal B}$(\Bzero\Tomumu) are not yet excluded by experimental data. The most stringent exclusion limit to date for this channel is about eight times larger than the SM prediction (at 95\% CL). The prospects for further improving the precision on the B$_{s,d}$ branching fractions to dimuons at the LHC are excellent. The LHCb, CMS and ATLAS experiments have collected about 2.1~fb$^{-1}$, 23~fb$^{-1}$ and 22~fb$^{-1}$ in 2012, respectively, at the center-of-mass energy $\sqrt{s}=8$~TeV. Results based on the analysis of these datasets are expected to be released within 2013.

%
%
\section{Studies of $b \rightarrow s \mu^+ \mu^-$ transitions\label{sec:b2smm}}
Rare B hadron decays of the type $b \rightarrow s \mu^+ \mu^-$ are mediated by FCNC and occur in the SM with branching ratios of the order O($10^{-6}$)~\cite{Melikhov:1997wp,Ali:1999mm}. Theories extending the SM favour enhanced rates of these decays and can modify the angular distributions of the decay products. Although no significant departures from the SM have been observed so far, the LHCb collaboration recently reported an isospin asymmetry between the branching ratio of $\mathrm{B^+} \rightarrow \mathrm{K^+}\mu^+\mu^-$ and $\mathrm{B^0} \rightarrow \mathrm{K^0}\mu^+\mu^-$ deviating from zero with a significance of 4.4 standard deviations~\cite{Aaij:2012cq}. In addition, the Belle experiment has measured a lepton forward-backward asymmetry ($A_\mathrm{FB}$) different from SM expectation at 2.7 standard deviations~\cite{Wei:2009zv}. 

\begin{figure}[tbp]
\begin{center}
\includegraphics[width=5.5cm]{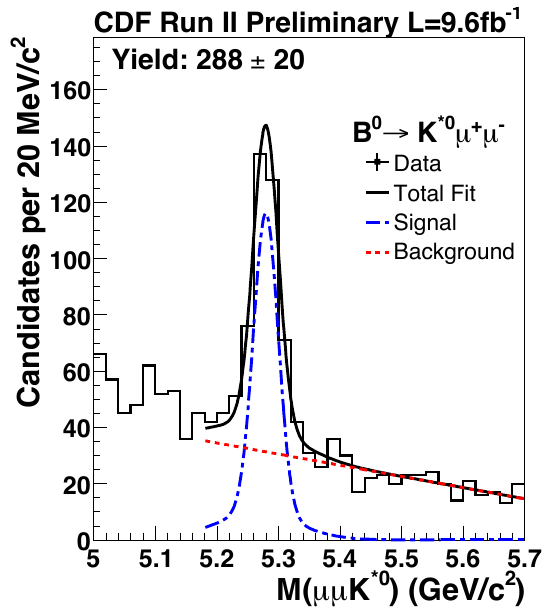}
\includegraphics[width=5.5cm]{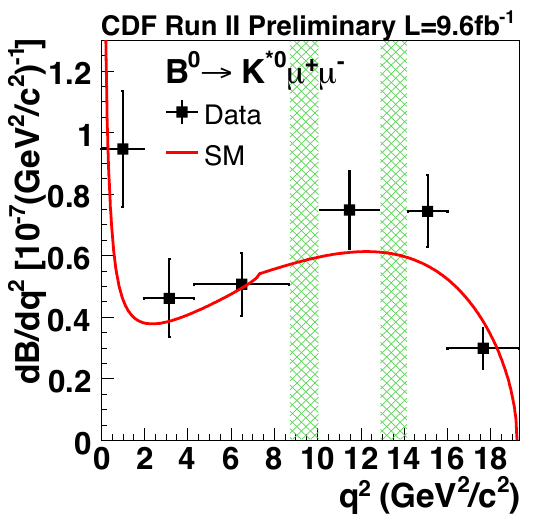}
\caption{{\it Up}: Invariant mass distribution for the $\mathrm{B^0} \rightarrow \mathrm{K^{*0}}\mu^+\mu^-$ channel measured by the CDF experiment. {\it Down}: Differential branching ratio as function of $q^2$  (squares) compared to the SM expectation (red line).}
\label{fig:bzeromass}
\end{center}
\end{figure}

The CDF experiment  recently analysed the full dataset corresponding to an integrated luminosity of 9.6 fb$^{-1}$. Besides the measurements of total and differential branching ratios in $\mathrm{B^+} \rightarrow \mathrm{K^+}\mu^+\mu^-$, $\mathrm{B^0} \rightarrow \mathrm{K^{*}}(892)^0\mu^+\mu^-$ , $\mathrm{B^0_s} \rightarrow \phi\mu^+\mu^-$ , $\mathrm{B^0} \rightarrow \mathrm{K^0}\mu^+\mu^-$ , $\mathrm{B^+} \rightarrow \mathrm{K^*}(892)^+\mu^+\mu^-$, and $\Lambda^0_b \rightarrow \Lambda \mu^+\mu^-$, measurements of the combined branching ratio assuming isospin symmetry and of the isospin asymmetry between neutral and charged B mesons are provided~\cite{CDFbtosmm}. To cancel the dominant systematic uncertainties, the decay rates for each rare channel $\mathrm{H_b} \rightarrow h\mu^+\mu^-$ were normalised to the corresponding resonant channel  $\mathrm{H_b} \rightarrow \mathrm{h} J/\psi$, where $\mathrm{H_b}$ can be B$^+$, B$^0$, B$^0_s$ and $\Lambda_b$, and  $h$ stands for K$^+$, K$^{*0}$, $\phi$, K$_S^0$, K$^{*+}$, and $\Lambda$. Candidate events were selected by constructing a vertex of two muons that satisfy the trigger requirements with one charged track or with two reconstructed tracks of opposite charge, for the case of charged and neutral $h$ hadrons, respectively. For the normalisation samples the dimuon invariant mass was required to be within 50 MeV of the $J/\psi$ mass. After loose selection cuts the rare decays were tightly selected with a multivariate analysis based on a Neural Network. The signal yields were obtained from an unbinned maximum likelihood fit to the invariant mass distribution, as shown in Figure~\ref{fig:bzeromass}, and were corrected for the MC selection efficiencies to extract the relative branching ratios. The results are summarised in Table~\ref{tab:bsmm_branchings}. The differential branching ratios were measured as function of the dimuon invariant mass, $q^2 = m^2_{\mu\mu} c^2$, by performing the signal fit in each $q^2$ bin, as shown in Figure~\ref{fig:bzeromass} for the $\mathrm{B^0} \rightarrow \mathrm{K^{*0}}\mu^+\mu^-$ decay. 

The isospin asymmetries between neutral and charged B meson decays are observables with relatively low theoretical uncertainties, since the leading form factor uncertainties cancel in the ratio. The asymmetries 
\[
A_I^{(*)} = \frac{\cal{B}(\mathrm{B^0} \rightarrow \mathrm{K^{(*)0}}\mu^+\mu^-) - \cal{B}(\mathrm{B^+} \rightarrow \mathrm{K^{(*)+}}\mu^+\mu^-)}{\cal{B}(\mathrm{B^0} \rightarrow \mathrm{K^{(*)0}}\mu^+\mu^-) + \cal{B}(\mathrm{B^+} \rightarrow \mathrm{K^{(*)+}}\mu^+\mu^-)}
\]
were also measured by the CDF experiment and were found to be consistent with zero over the full $q^2$ range. Thus this result does not confirm the deviation observed by LHCb.

\begin{figure*}[htbp]
\begin{center}
\includegraphics[width=10cm]{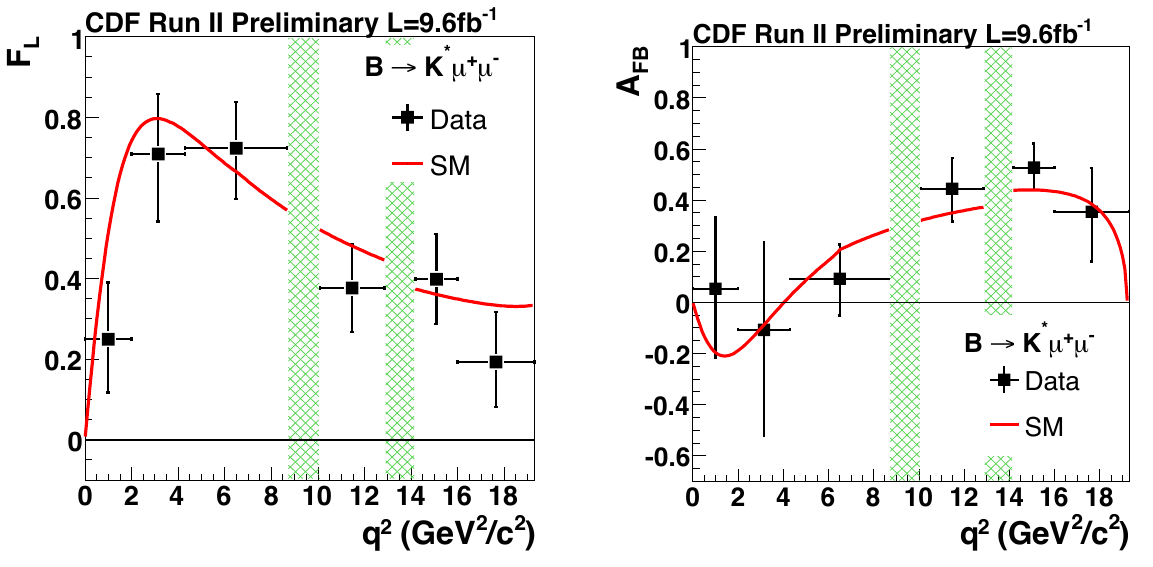}
\caption{Longitudinal polarisation (left) and forward-backward asymmetry (right) for $\mathrm{B^0} \rightarrow \mathrm{K^{*}}\mu^+\mu^-$ decays as function of the dimuon invariant mass. The SM expectation is represented by the solid line.}
\label{fig:anganalysis}
\end{center}
\end{figure*}

In $\mathrm{B^0}\rightarrow \mathrm{K^{(*)}}\mu^+\mu^-$ decays the muon forward-backward asymmetry ($A_\mathrm{FB}$) and the $\mathrm{K^*}$ longitudinal polarisation ($F_{\mathrm L}$) are extracted from the angular distributions $\cos \theta_\mu$ and $\cos \theta_K$, respectively, where $\theta_\mu$ is the angle between the $\mu^+$ direction and the opposite of the B meson direction in the dimuon rest frame, while $\theta_K$ is the angle between the kaon direction and the direction opposite to the B meson in the $\mathrm{K^*}$ rest frame. A simultaneous unbinned maximum likelihood fit of the polarisation angles is performed to extract the angular variables as function of the dimuon invariant mass, as shown in Figure~\ref{fig:anganalysis}. All measurements are consistent with the SM prediction and with previous measurements.
\begin{table*}[htdp]
\caption{Relative branching ratios for $b \rightarrow s \mu^+\mu^-$ decays measured with the CDF experiment.}
\begin{center}
\begin{tabular}{c|c}
Channel ratio & Relative branching ratio ($10^{-3}$) \\ \hline
${\cal B}(\mathrm{B^+} \rightarrow \mathrm{K^+}\mu^+\mu^-)/{\cal B}(\mathrm{B^+} \rightarrow \mathrm{K^+}J/\psi)$ & $0.44 \pm 0.03$(stat)$ \pm 0.02$(syst) \\
${\cal B}(\mathrm{B^0} \rightarrow \mathrm{K^{*0}}\mu^+\mu^-)/{\cal B}(\mathrm{B^0} \rightarrow \mathrm{K^{*0}}J/\psi)$ & $0.85 \pm 0.07$(stat)$ \pm 0.03$(syst) \\
${\cal B}(\mathrm{B^0_s} \rightarrow \phi\mu^+\mu^-)/{\cal B}(\mathrm{B^0_s} \rightarrow\phi J/\psi)$ & $0.90 \pm 0.14$(stat)$ \pm 0.07$(syst) \\
${\cal B}(\mathrm{B^0} \rightarrow \mathrm{K^{0}}\mu^+\mu^-)/{\cal B}(\mathrm{B^0} \rightarrow \mathrm{K^{0}}J/\psi)$ & $0.38 \pm 0.10$(stat)$ \pm 0.03$(syst) \\
${\cal B}(\mathrm{B^+} \rightarrow \mathrm{K^{*+}}\mu^+\mu^-)/{\cal B}(\mathrm{B^+} \rightarrow \mathrm{K^{*+}}J/\psi)$ & $0.62 \pm 0.18$(stat)$ \pm 0.06$(syst) \\
${\cal B}(\Lambda_b^0 \rightarrow \Lambda\mu^+\mu^-)/{\cal B}(\Lambda_b^0 \rightarrow \Lambda J/\psi)$ & $2.75 \pm 0.48$(stat)$ \pm 0.27$(syst) \\
\end{tabular}
\end{center}
\label{tab:bsmm_branchings}
\end{table*}%

%

\end{document}